\documentclass[conference]{IEEEtran}
\IEEEoverridecommandlockouts
\usepackage{cite}
\usepackage{amsmath,amssymb,amsfonts}
\usepackage{algorithmic}
\usepackage{graphicx}
\usepackage{textcomp}
\usepackage{xcolor}
\usepackage{balance}
\usepackage{hyperref}
\usepackage{caption,subcaption}
\def\BibTeX{{\rm B\kern-.05em{\sc i\kern-.025em b}\kern-.08em  T\kern-.1667em\lower.7ex\hbox{E}\kern-.125emX}}
\begin{document}

\title{Optimal Stochastic Resource Allocation for Distributed Quantum Computing}
\author{\IEEEauthorblockN{Napat Ngoenriang\IEEEauthorrefmark{1},
Minrui Xu\IEEEauthorrefmark{2}, Sucha Supittayapornpong\IEEEauthorrefmark{1},
Dusit Niyato\IEEEauthorrefmark{2}, Han Yu\IEEEauthorrefmark{2}, Xuemin (Sherman) Shen\IEEEauthorrefmark{3}}\\
\IEEEauthorblockA{\IEEEauthorrefmark{1}School of Information Science and Technology, Vidyasirimedhi Institute of Science and Technology, Thailand\\
\IEEEauthorrefmark{2}School of Computer Science and Engineering, Nanyang Technological University, Singapore\\
\IEEEauthorrefmark{3}Department of Electrical and Computer Engineering, University of Waterloo, Waterloo, ON, Canada
}}

\maketitle
\begin{abstract}
With the advent of interconnected quantum computers, i.e., distributed quantum computing (DQC), multiple quantum computers can now collaborate via quantum networks to perform massively complex computational tasks. However, DQC faces problems sharing quantum information because it cannot be cloned or duplicated between quantum computers. Thanks to advanced quantum mechanics, quantum computers can teleport quantum information across quantum networks. However, challenges to utilizing efficiently quantum resources, e.g., quantum computers and quantum channels, arise in DQC due to their capabilities and properties, such as uncertain qubit fidelity and quantum channel noise. In this paper, we propose a resource allocation scheme for DQC based on stochastic programming to minimize the total deployment cost for quantum resources. Essentially, the two-stage stochastic programming model is formulated to handle the uncertainty of quantum computing demands, computing power, and fidelity in quantum networks. The performance evaluation demonstrates the effectiveness and ability of the proposed scheme to balance the utilization of quantum computers and on-demand quantum computers while minimizing the overall cost of provisioning under uncertainty.
\end{abstract}

\begin{IEEEkeywords}
Distributed Quantum Computing, Quantum Networks, Resource Allocation, Stochastic Programming
\end{IEEEkeywords}

\section{Introduction}
To date, quantum computing is tremendously fast and efficient, being able to do computations in a matter of seconds which would take decades for older supercomputers~\cite{qibm}. 
In 2019, a prototype of Google's quantum computer was able to finish the computation and demonstrate the effectiveness of quantum mechanics~\cite{qai}. Breakthroughs in quantum computing are essential and influence various applications, such as artificial intelligence (AI), molecular modelling, weather forecasting, and drug development~\cite{qmarket}.
Since the development of quantum computers is still in its infancy, distributed quantum computing (DQC) has emerged in significance to solve more complex computational tasks. 
In the next quantum computing era, IBM~\cite{qibmdqc} and Google~\cite{qgoogledqc} aim to introduce a practical DQC, which is anticipated in 2025. 
Although the advent of DQC has also advanced processing speed for a variety of heterogeneous tasks, quantum resources are still limited and need to be managed effectively while performing computations.

\begin{figure*}[t]
    \centering \includegraphics[width=\linewidth]{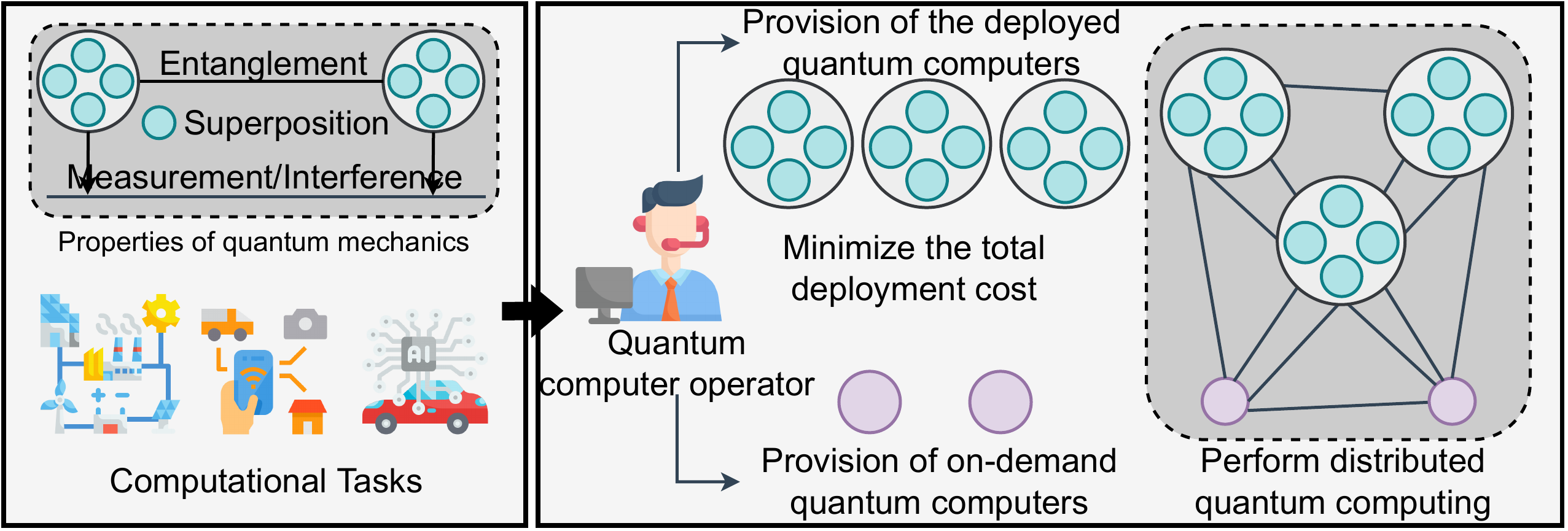}
    \caption{The illustration of system model of distributed quantum computing.}
    \label{fig:diagram}
\end{figure*}

The basis for the development of quantum computing is formed based on properties of quantum mechanics, i.e., superposition, entanglement, and interference, as shown in Fig.~\ref{fig:diagram}. \textit{Superposition} permits encoding in a mixture of two states (i.e., qubits). A qubit offers more information storage and computation choices than binary bits in classical computers. In quantum mechanics, \textit{entanglement} describes a correlation between two qubits in which the values of one qubit may depend on another. 
To observe the values of qubits, measurement, often referred to as \textit{interference}, can be used to interrupt the processing of quantum computers. 

Quantum algorithms enable qubits in quantum computers by utilizing quantum mechanics. For instance, Shor's~\cite{shor1999polynomial} and Grover's~\cite{grover1997quantum} algorithms were developed to deal with factorization and the search for unstructured data, respectively, which are highly challenging for classical computers. To tackle these tasks with quantum computers, at least the order of $10^6$ physical qubits are required~\cite{shor1999polynomial}. However, a recently developed quantum computer can only contain tens of qubits~\cite{arute2019quantum}. Therefore, it becomes increasingly challenging to handle and control information in quantum computers due to the limited number of qubits, the instability of qubits, and the amount of information required for complex computational tasks. As a result, the concept of DQC has been presented.

In DQC, quantum teleportation, or the transfer of qubits, is required for quantum computers to connect and collaborate. 
In this regard, multiple quantum computers can work collaboratively to compute a large-scale complex computational task at quadratic or exponential speed-up. 
Moreover, most of common quantum algorithms can benefit from their distributed equivalents. For example, distributed Shor’s algorithm can reduce computational complexity compared to original Shor’s algorithm~\cite{yimsiriwattana2004distributed}. Additionally, distributed Grover’s algorithm has a considerably lower query time than Grover’s algorithm~\cite{qiu2022distributed}. Therefore, distributed quantum algorithms can enhance practical feasibility of quantum computers in tackling complex computational tasks in practice.

Although DQC and distributed quantum algorithms have evolved and accelerated to handle complex computational tasks, quantum resources, e.g., quantum computers and quantum channels, must be optimally allocated to perform heterogeneous computational tasks. However, the efficient utilization of quantum resources in DQC is still facing challenges.
First, the utilization of quantum resources depends on the demands of computational tasks, which are not known precisely at the time of quantum computer deployment. 
Second, quantum computers' availability and computing power may or may not be available to compute computational tasks. 
Third, fidelity degradation occurs when performing quantum teleportation and transferring qubits over quantum networks, which degrades the efficacy of the entangled qubits.
Due to the uncertainty, allocating quantum resources in DQC may result in both under and over-deployment of quantum computers. 
In addition, because of the limitations of quantum resources, quantum computers can be purchased or outsourced from other organizations such as Amazon Braket~\cite{qcloud} to complete complex computational tasks. However, the cost of on-demand quantum computers is relatively more expensive.

To address the aforementioned challenges, in this paper, we propose a resource allocation scheme based on stochastic programming to minimize the total deployment cost to compute computational tasks in DQC.
The main contributions of this paper can be summarized as follows:
\begin{itemize}
    \item We propose an optimal resource allocation scheme to compute computational tasks in DQC. In particular, the optimal resource allocation are jointly obtained under the uncertainties of future demands of computational tasks and instability of quantum characteristics.
    \item We conduct extensive experiments to demonstrate the importance and effectiveness of the proposed optimal resource allocation scheme, which achieves the lowest total deployment cost. 
\end{itemize}  

\section{Related Work}


Quantum computing was developed to increase the capability of existing computational resources~\cite{quantumbook1}.
One of challenges in quantum computing is quantum algorithms implemented on quantum computers, for example, Shor's~\cite{shor1999polynomial} and Grover's~\cite{grover1997quantum} algorithms are requiring a massive number of qubits to execute. For instance, Shor's algorithm was developed to handle a factorization problem with around $10^6$ qubits, which is too complex for classical computers~\cite{shor1999polynomial}. In addition, Grover's algorithm was presented to search unordered data by encoding inputs with dimension $N$ as superposition with $\sqrt{N}$ qubits, which gives a quadratic computing speed-up~\cite{grover1997quantum} to search operations in non-structure databases. These quantum algorithms can be adopted to outperform existing conventional algorithms with quadratic or exponential  speed-up computations~\cite{raj2018analysis}. 

However, due to challenges in scaling up the number of qubits, quantum computers are not yet ready to replace traditional computers~\cite{wehner2018quantum}. 
Therefore, DQC was introduced to combine several quantum computers to handle more complex computational tasks~\cite{wehner2018quantum,8910635}.
Most existing works look at distributed quantum mechanics and algorithms to be the basis in DQC~\cite{wehner2018quantum,8910635}. 
Similar to quantum algorithms, distributed Shor's~\cite{yimsiriwattana2004distributed} and Grover's~\cite{qiu2022distributed} algorithms were widely used.
The distributed Shor's algorithm has a computing complexity of $O((log N)^2)$, which outperforms the original Shor's algorithm, i.e., $O((log N)^2 log (log N) log (log (log N)))$, where $N$ is the number to be factored \cite{yimsiriwattana2004distributed}.
Grover's algorithm incurs expensive query time due to its use of superpositions as inputs. According to~\cite{yimsiriwattana2004distributed}, the distributed Grover's algorithm was devised to improve the original Grover's algorithm by reducing the query times from $\frac{\pi}{4} \sqrt{2^{n}}$ to $\frac{\pi}{4} \sqrt{2^{n-k}}$, where $n$ is the number of input bits of the original problem, whereas $n - k$ is the number of input bits for the decomposed subfunctions.

Next, we discuss resource allocation problems in distributed computing, quantum computing, and DQC, which is the motivation for our work, as follows:


\begin{itemize}
    \item[1)] \textit{Resource Allocation in Distributed Computing:} 
    To enhance the performance of distributed computing, the authors~\cite{haji2020dynamic} reviewed existing resource allocation schemes under dynamic environments in distributed computing with various types of classical resources, e.g., computing, power, and storage resources. In particular, the authors in~\cite{chaisiri2011optimization} formulated the problem of resource allocation in cloud computing as a stochastic programming model by considering the uncertainty of user requirements.

    \item[2)] \textit{Resource Allocation in Quantum Computing:}
    The authors in~\cite{ravi2021quantum,ravi2021adaptive} proposed and analyzed the significance of resource allocation problems and adaptive resource allocation problems, respectively, in quantum cloud computing, using quantum and cloud characteristics such as execution times, cloud query times, and circuit compilation times. 

    \item[3)] \textit{Resource Allocation in Distributed Quantum Computing:} In~\cite{9821091}, the authors proposed network flow optimization for DQC. The authors used a weighted round-robin algorithm to pre-compute traffic flows for all the possible paths for each application and then allocate resources to the applications in the round-robin, where the maximum net rate of the application is a ratio of the round size proportional to its weight.
\end{itemize}

However, all the existing studies overlook the issue of quantum resource allocation, e.g., quantum computers and channels, in addition to the uncertainty of quantum computing demands, computational power, and fidelity in DQC, which directly affects the use of quantum resources in DQC.

\section{System Model}
\subsection{System Overview}
We consider the system model of a quantum computer operator that provisions a quantum computing task by using quantum computers as shown in Fig. \ref{fig:diagram}. There are two options for the quantum computer operator to compute computational tasks on DQC, i.e., using the deployed quantum computers or using on-demand quantum computers from other organizations such as Amazon Braket~\cite{qcloud}.
Let $\mathcal{J}=\{1,\dots,j,\dots,J\}$ be a set of quantum computers. 
The number of quantum bits, or qubits (a quantum computation unit), is required to complete the computational task, denoted by $n$.  
In addition, $n$ qubits mean that $n$ classical bits can represent up to $2^n$ different possibilities~\cite{quantumbook1}.
In detail, we consider that quantum computer $j$ owns $k_{j}$ qubits. 
Alternatively, on-demand quantum computers can be purchased from other organizations and then connected across the deployed quantum computers to compute computational tasks. Let $\mathcal{R}=\{1,\dots,r,\dots,R\}$ represent a set of on-demand quantum computers.

\begin{figure}[t]
    \centering \includegraphics[width=.9\linewidth]{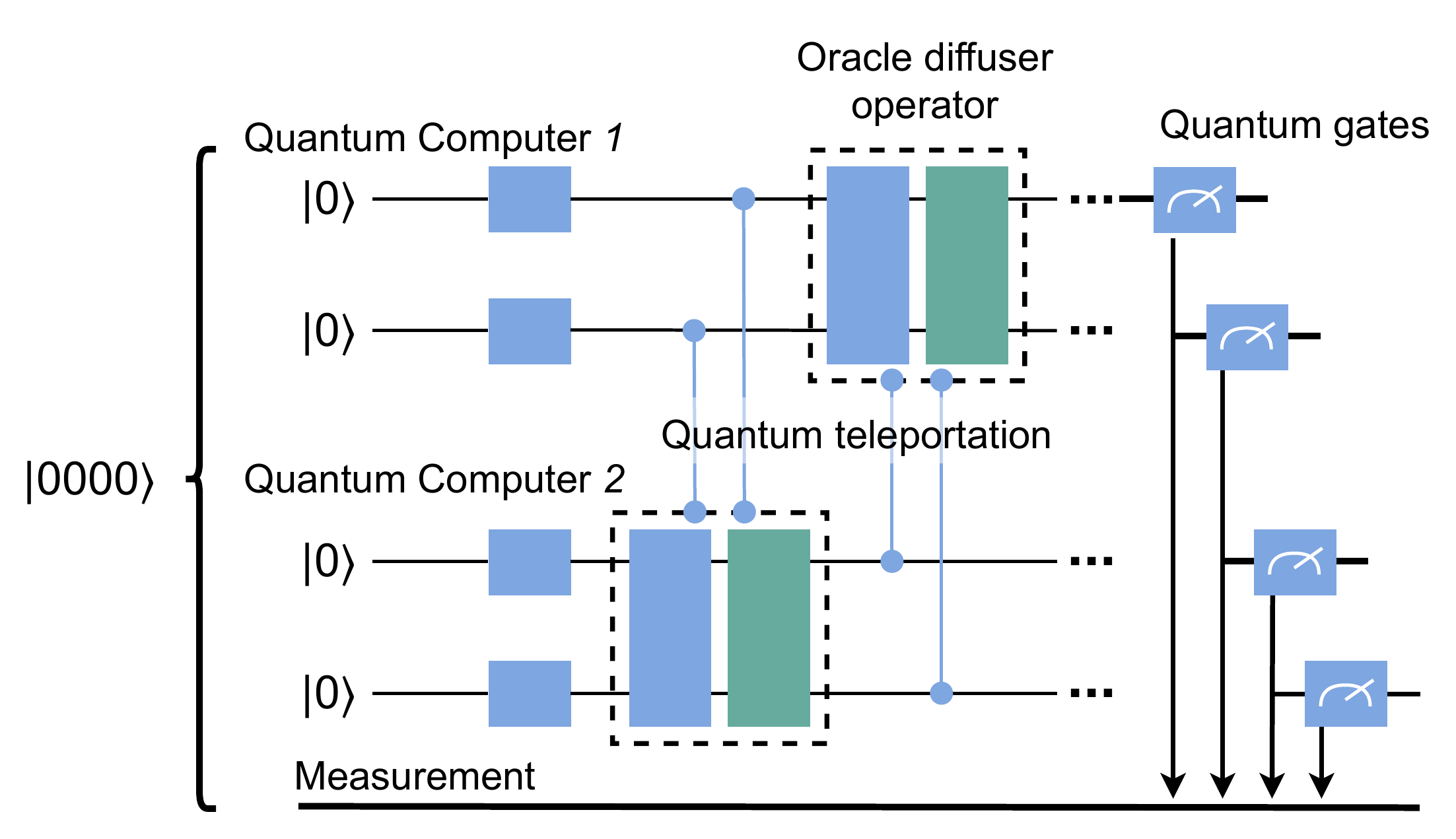}
    \caption{A procedure of distributed quantum computing of two quantum computers with quantum teleportation.}
    \label{fig:quante}
\end{figure}

\subsection{Quantum Teleportation Across Quantum Networks}
Due to the no-cloning theorem in quantum mechanics, qubits cannot be duplicated or cloned across quantum computers~\cite{wehner2018quantum}. The concept of quantum teleportation enables two quantum computers to exchange qubits, as shown in Fig. \ref{fig:quante}. The quantum teleportation transfers a pair of qubits executing from the source to the destination quantum computers to local operations and then measures the qubits previous to the qubits' decoherence (i.e., loss of information in qubits due to the instability of quantum characteristics). Consequently, multiple quantum computers can connect and collaborate to complete computational tasks. 
The quantum computers interconnected via links are directional with a fixed capacity, indicated by $C_{i,j} > 0,$ where $i$ and $j$ are two connected quantum computers. However, errors may occur while preserving the entangled qubits between quantum computers. In particular, the fidelity of the shared Bell pair, indicated by $q_{i,j}$, is a performance that measures the effectiveness of the entanglement between the desired and actual states of the quantum teleportation~\cite{jozsa1994fidelity}. The range of fidelity is $[0, 1]$, where $1$ refers to the best performance that the maximally entangled qubits can be reached~\cite{9821091}.

\subsection{Uncertainty}
We classified the uncertainty in DQC into three types, i.e., i) the demand of the computational task, ii) the availability and computing power of the quantum computers, and iii) the fidelity of the entangled qubits. Let $\Omega = \{\omega_1, \dots, \omega_{|\Omega|}\}$ denote the set of scenarios that describes the demand of computational tasks, the computing power of quantum computers, and the fidelity of the Bell pairs, where $|\Omega|$ is the number of total scenarios. Let $\pi(\omega)$ be the probability that the scenario $\omega \in \Omega$ is realized, where $\pi(\omega)$ can be calculated based on historical data~\cite{chaisiri2011optimization}.

\begin{itemize}
    \item[i)] The actual demands of computational tasks are unknown at the time of deploying the quantum computer. Different computational tasks such as minimization, material science, and machine learning problems may require various qubits~\cite{qapp}. For example, 58 qubits are required in material science problems~\cite{matqubits}. Let $\Tilde{n}(\omega)$ be an integer parameter indicating the demand of computational tasks. 
    \item[ii)] The precise availability and computing power of the quantum computer are unknown since it could be set aside for other applications or because its backend might not support all of them~\cite{ibmql}. 
    Let $\Tilde{k}_{j} (\omega)$ be an integer parameter indicating the computing power of the quantum computer $j$ in qubits. 
    \item[iii)] Specifically, the fidelity of the shared entangled qubits, also known as the Bell pair, in DQC is also not known exactly~\cite{jozsa1994fidelity}. 
    Let $\Tilde{q}_{i,j}(\omega)$ denote the fidelity of the Bell pair between quantum computers $i$ and $j$.
\end{itemize}

For example, $\Tilde{n}(\omega) = 1, \Tilde{k}_{1} (\omega) = 2, $ and $\Tilde{q}_{1,2}(\omega) = 0.5$ mean that the demand of the computational task is 1 qubit (or $2^1$ in bit), the computing power of the quantum computer $1$ is 2 qubits, and the fidelity of the Bell pair for quantum teleportation between the quantum computers $1$ and $2$ is 0.5.

\subsection{Cost}
One of the main obstacles to the practical deployment of DQC is the increased costs of using quantum computers, requiring computing power, and using the Bell pair for the entangled qubits~\cite{9684555}.
Four associated costs of quantum computing are described. $c^{(dep)}_{j}$ represents the deployment cost of the quantum computer $j$. $c^{(com)}_{j}$ represents the unit cost of the computing power of the quantum computer $j$ in qubits. $c^{(Bell)}_{i,j}$ represents the cost of the Bell pair for the shared entangled qubits of the quantum computers $i$ and $j$. $c^{(ond)}_{r}$ represents the additional cost associated with the deployment of an on-demand quantum computer $r$.

\section{Problem Formulation}
This section presents the deterministic integer programming and stochastic integer programming formulations to minimize the total deployment cost for the quantum computer operator.

\subsection{Deterministic Integer Programming}
If the demand of the computational task, computing power of quantum computers, and fidelity of the Bell pairs are completely known, quantum computers can certainly be deployed. Therefore, an on-demand quantum computer is not necessary. This problem can be formulated as deterministic integer non-linear programming~\cite{birge2011introduction}. A decision variable is listed below.
\begin{itemize}
    \item $x_{j}$ indicates whether the quantum computer $j$ is reserved to compute the computational task.
\end{itemize}
The deterministic formulation is defined as follows:
\begin{align}
    \operatorname*{\min_{x_{j}}:} \sum_{j=1}^{J} c^{(dep)}_{j} x_{j} + \sum_{j=1}^{J} c^{(com)}_{j} x_{j} + \sum_{j=1}^{J} c^{(Bell)}_{i,j} x_{i}x_{j}, \label{obj:det}
\end{align}
subject to
\begin{align}
    &\sum_{j=1}^{J} k_{j} x_{j} \geq 2^{n}, & & \label{constr:finishtaskdet}\\
    &C_{i,j}q_{i,j} \geq \min\{x_{i}k_{i}, x_{j}k_{j}\},&  \forall i,j \in \mathcal{J}, i \ne j,& \label{constr:linkcapdet}\\
    & x_{j} \in \{0,1\},&  \forall j \in \mathcal{J}.&\label{constr:binarydet}
\end{align}

The objective in \eqref{obj:det} is to minimize the total deployment cost for the quantum computer operator that depends on the deployment of the quantum computers. The constraint in \eqref{constr:finishtaskdet} enforces that the computing power of the used quantum computers must meet the demand of the computational task in $n$ qubits indicating $2^n$ bits. The constraint in \eqref{constr:linkcapdet} ensures that the entanglement of qubits can be successful if the computing power of the used quantum computers does not exceed the link capacity~\cite{9821091}. The constraint in \eqref{constr:binarydet} indicates that $x_{j}$ is a binary variable.

\subsection{Stochastic Integer Programming}
In reality, the demands of the computational task, computing power, and fidelity of the Bell pairs cannot be predicted at the time of quantum computer deployment. Thus, deterministic integer programming is unsuitable. Therefore, we formulate a two-stage stochastic non-linear integer programming~\cite{birge2011introduction}. The deployment of quantum computers is defined in the first stage, and their utilization and on-demand quantum computers are defined in the second stage. The first-stage and second-stage decisions are made before and after observing the actual demand, computing power of the quantum computers, and the fidelity of the Bell pairs, respectively.
Decision variables are listed as follows:
\begin{itemize}
    \item $\Tilde{x}_{j} (\omega)$ indicates whether the deployed quantum computer $j$ is used to compute the computational task.
    \item $y_{r} (\omega)$ indicates whether the on-demand quantum computer $r$ is deployed to compute the computational task. 
\end{itemize}
The stochastic formulation is defined as follows:
\begin{align}
    \operatorname*{\min_{x_{j},\Tilde{x}_{j} (\omega), y_{r}(\omega)}:} \sum_{j=1}^{J} c^{(dep)}_{j} x_{j} + \mathbb{E} [\mathcal{Q} (\Tilde{x}_{j}, y_{r}, \omega)],
\end{align}
\begin{align}
\mathbb{E}& [\mathcal{Q} (y_{r}, \omega)] = \sum_{\omega \in \Omega} \pi(\omega) \bigg( \sum_{j=1}^{J} c^{(com)}_{j} \Tilde{x}_{j} (\omega) \nonumber\\ &+ \sum_{j=1}^{J} \sum_{i=1}^{J\backslash \{j\}}  c^{(Bell)}_{i,j} \Tilde{x}_{i} (\omega) \Tilde{x}_{j} (\omega) + \sum_{r=1}^{R} c^{(ond)}_{r} y_{r}(\omega) \bigg), \label{obj:sto}
\end{align}
subject to
\begin{align}
    &\Tilde{x}_{j} (\omega) \leq x_{j}, &  \hspace{-8em} \forall j \in \mathcal{J}, \forall \omega \in \Omega,& \label{constr:usednodeinsto}\\
    &\sum_{j=1}^{J} \Tilde{k}_{j} (\omega) \Tilde{x}_{j} (\omega) + \sum_{r=1}^{R} g_{r} y_{r} (\omega)  \geq 2^{\Tilde{n}(\omega)}, &  \hspace{-8em} \forall \omega \in \Omega,& \label{constr:finishtasksto}\\
    &C_{i,j}\Tilde{q}_{i,j}(\omega) \geq \min\{\Tilde{k}_{i} (\omega) \Tilde{x}_{i} (\omega), \Tilde{k}_{j} (\omega) \Tilde{x}_{j} (\omega)\},& &
    \nonumber\\ & & \hspace{-8em} \forall i,j \in \mathcal{J}, i \ne j, \forall \omega \in \Omega,  &\label{constr:linkcapsto}\\
    &\Tilde{x}_{j} (\omega), y_{j} (\omega) \in \{0,1\},&  \hspace{-8em}\forall j \in \mathcal{J}, \forall \omega \in \Omega \label{constr:binarysto}.&
\end{align}

The objective in \eqref{obj:sto} is to minimize the total deployment cost for the quantum computer operator under uncertainty. The constraint in \eqref{constr:usednodeinsto} ensures that the utilized quantum computer for computing the computational task does not exceed the selected quantum computer. The constraint in \eqref{constr:finishtasksto} enforces that the computing power of both the utilized and on-demand quantum computers can complete the computational task requiring $n$ qubits, where $g_{r}$ indicates the computing power of the on-demand quantum computer. The constraint in \eqref{constr:linkcapsto} ensures that the computing power of the quantum computers does not exceed the capability of the link capacity. The constraint in \eqref{constr:binarysto} indicates that $\Tilde{x}_{j} (\omega)$ and $ y_{j} (\omega) $ are a binary variable.

In addition, the stochastic non-linear programming can be reformulated as a deterministic non-linear equivalent formulation~\cite{birge2011introduction}, i.e., mixed-integer non-linear programming (MINLP), which can be solved by conventional optimization solver software using synergies with mixed-integer programming and non-linear programming methods~\cite{gams}.

\section{Performance Evaluation}
\begin{figure*}[ht]
\begin{minipage}{0.33\linewidth}
\includegraphics[width=\textwidth]{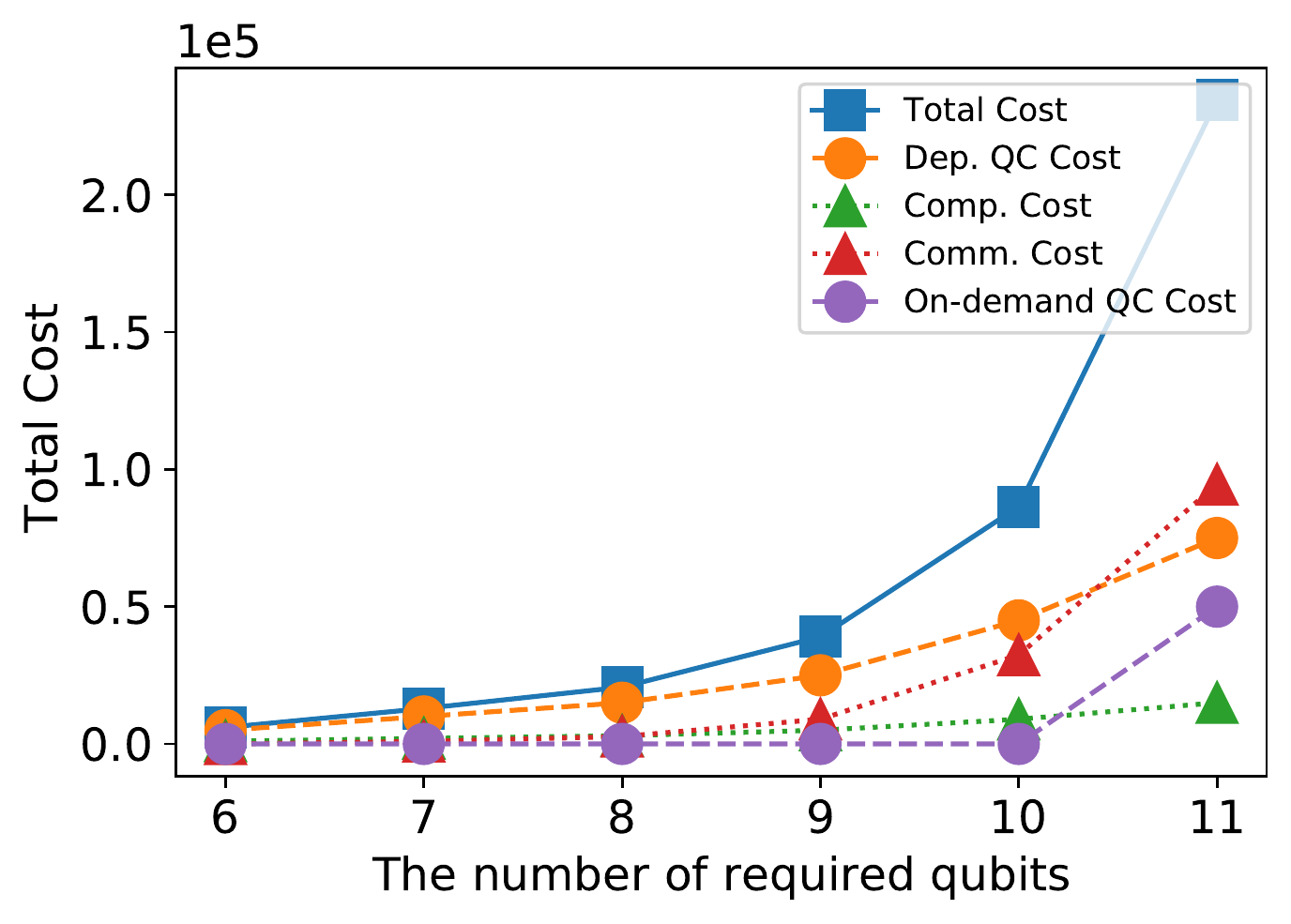}
\caption{Cost breakdown under different demands of the computational task.}
\label{fig:demand}
\end{minipage}%
\hfill
\begin{minipage}{0.33\linewidth}
\includegraphics[width=\textwidth]{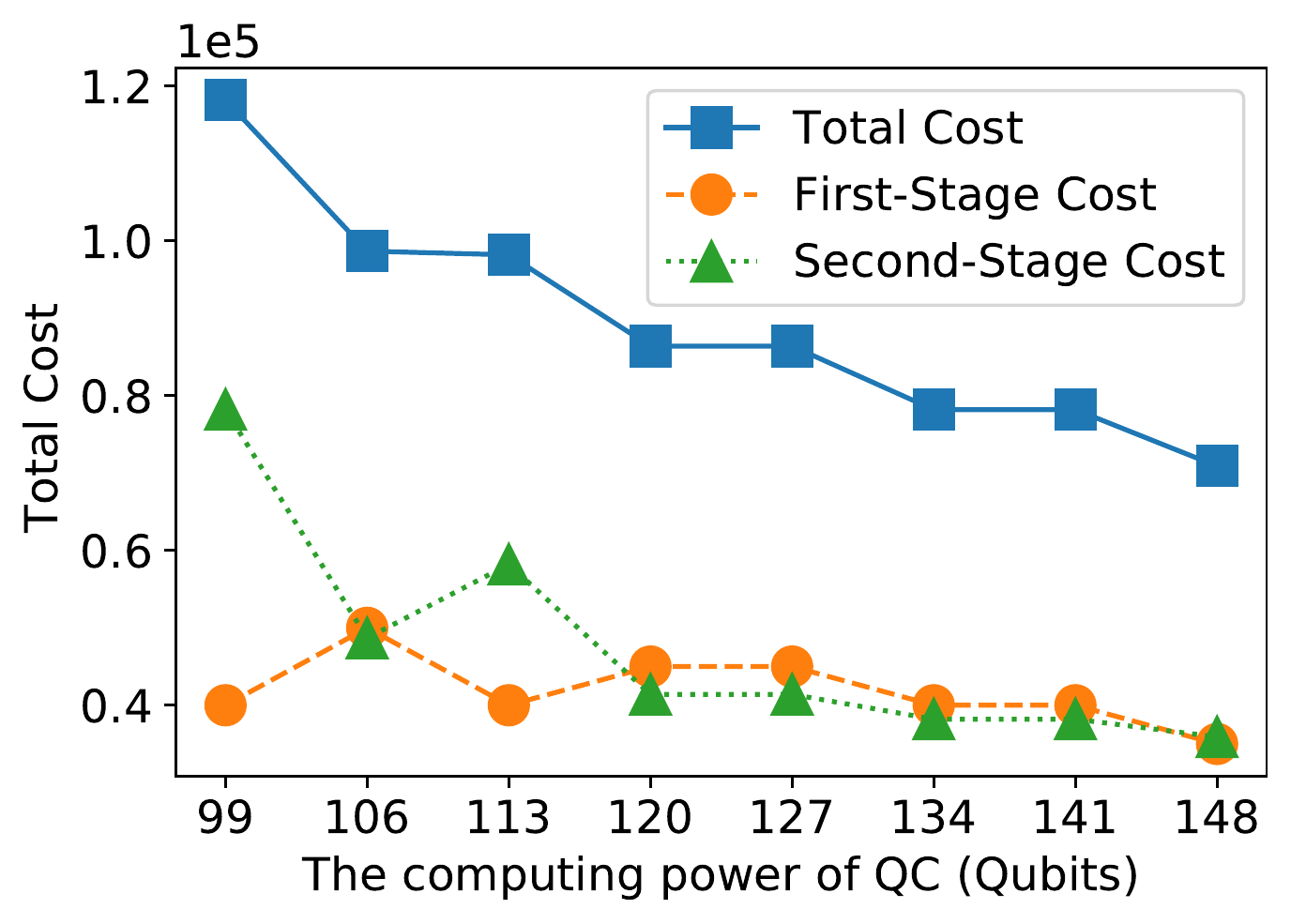}
\caption{Cost structure under different computing power of the quantum computer.}
\label{fig:cap}
\end{minipage}%
\hfill
\begin{minipage}{0.33\linewidth}
\includegraphics[width=\textwidth]{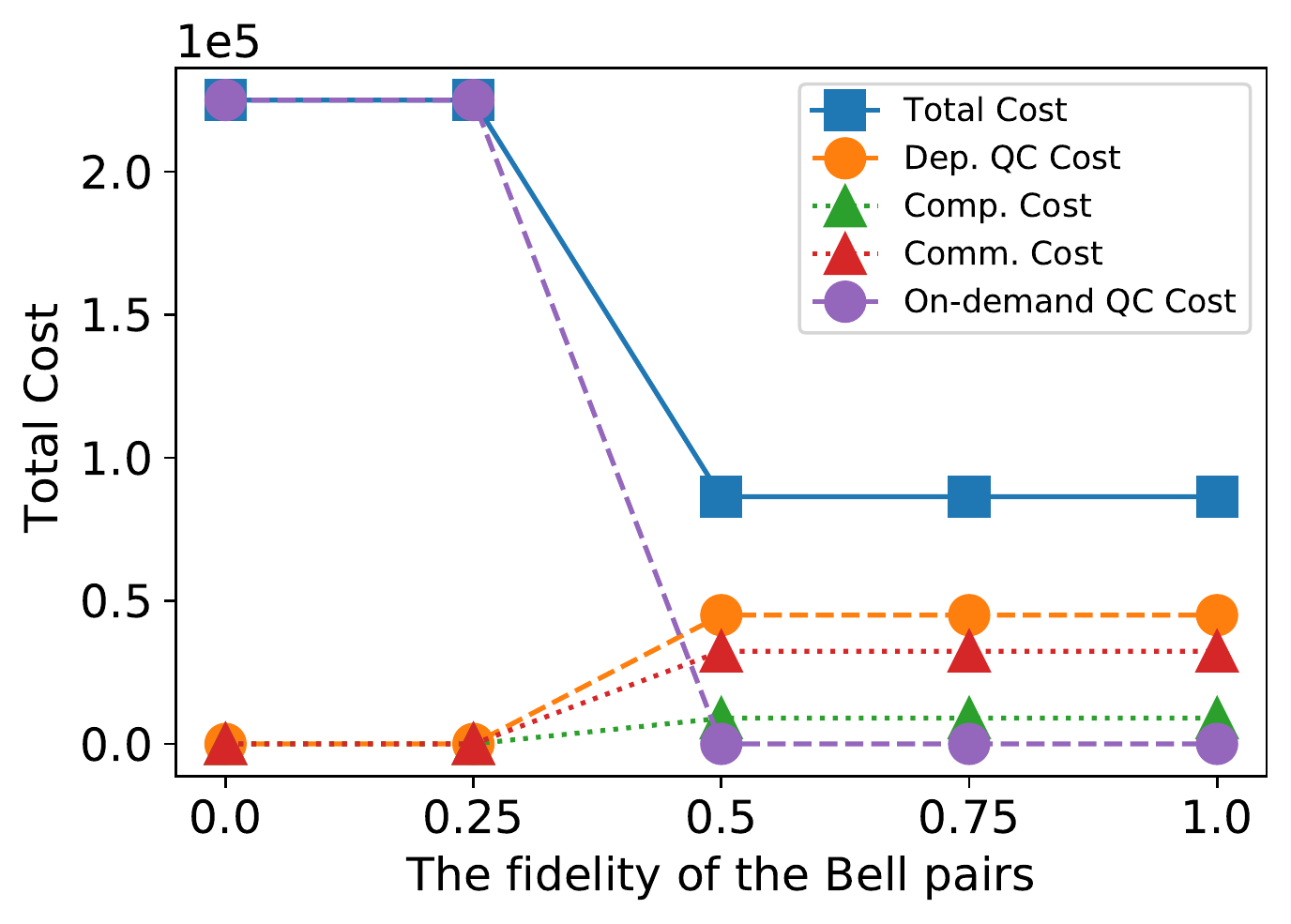}
\caption{Cost breakdown under different fidelity of the Bell pairs.}
\label{fig:fid}
\end{minipage}
\end{figure*}

\begin{figure*}[ht]
\begin{minipage}{0.33\linewidth}
\includegraphics[width=\textwidth]{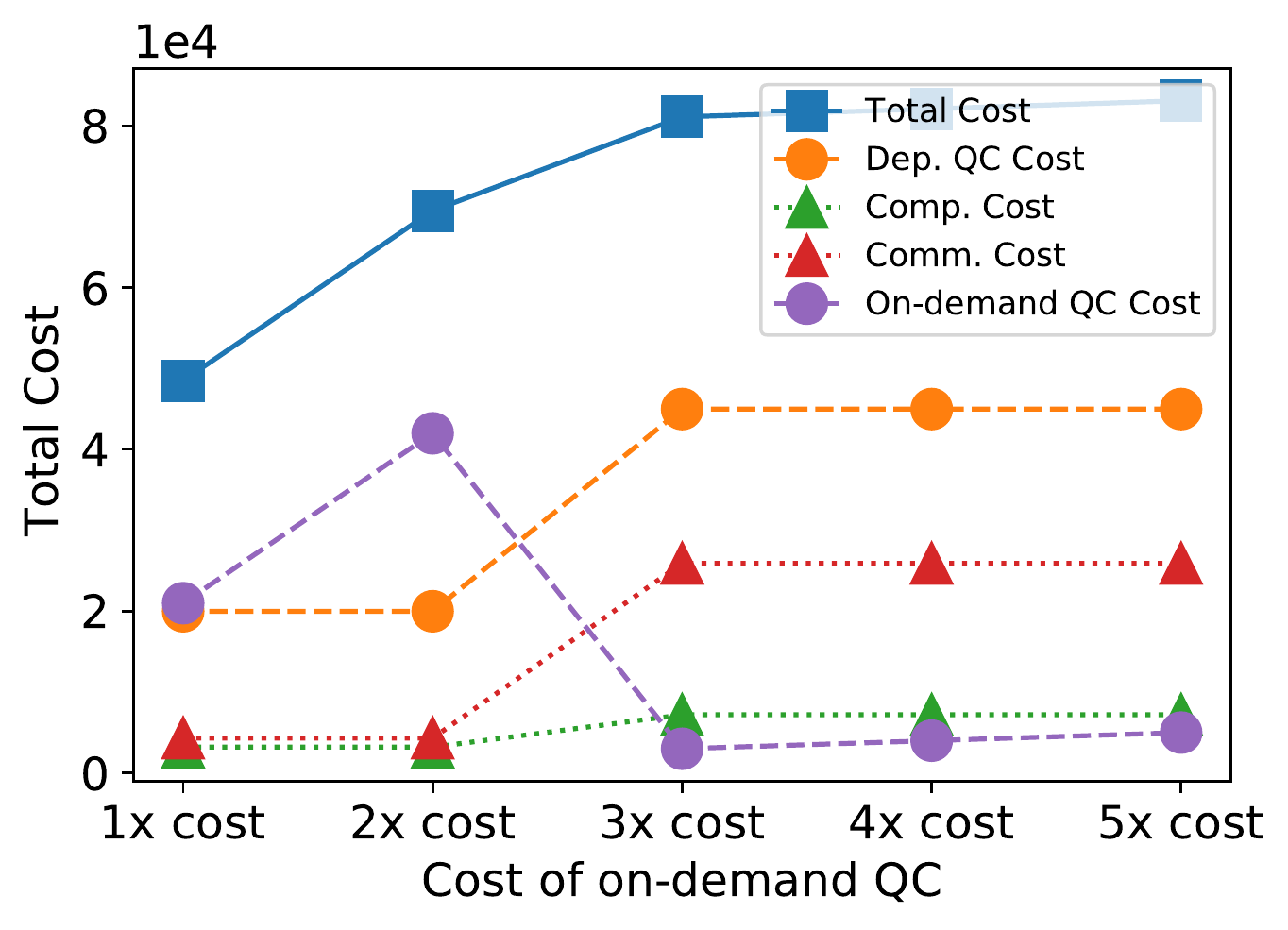}
\caption{Cost breakdown under different on-demand QC costs.}
\label{fig:newqc}
\end{minipage}%
\hfill
\begin{minipage}{0.33\linewidth}
\includegraphics[width=\textwidth]{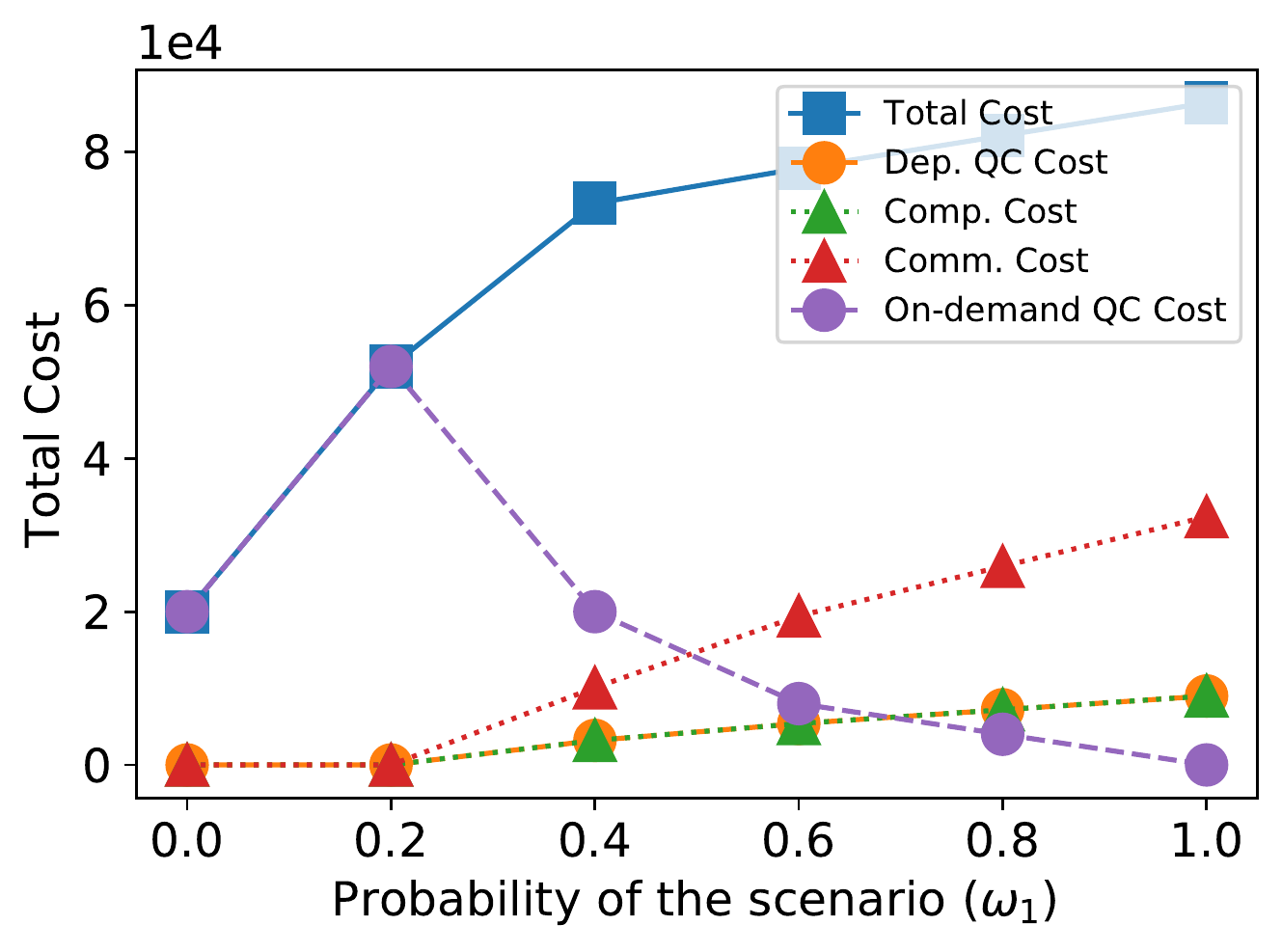}
\caption{Cost breakdown under different probabilities.}
\label{fig:prob}
\end{minipage}%
\hfill
\begin{minipage}{0.33\linewidth}
\includegraphics[width=\textwidth]{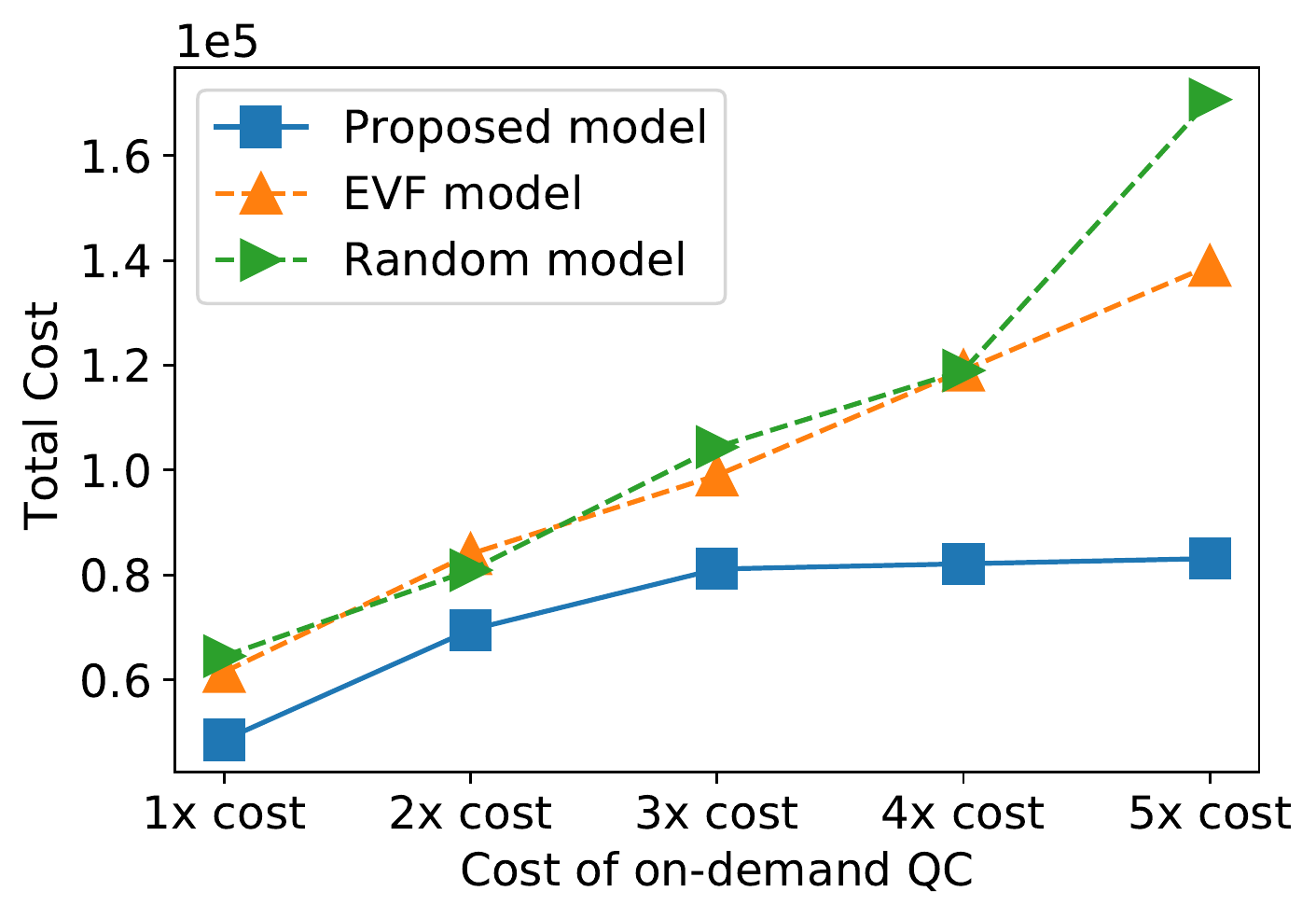}
\caption{Cost comparison among the proposed, EVF, and random models.}
\label{fig:compare}
\end{minipage}
\end{figure*}

\subsection{Parameter Setting}
We consider the system model of DQC, where the quantum computer operator consists of 10 quantum computers. We set the cost values, measured in normalized monetary, for the deployment of quantum computers, computing power, and Bell pairs for the shared entangled qubits to be 5000, 1000, and 450, respectively, based on~\cite{9684555}. All Bell pairs among quantum computers have identical costs and qubit capacities, i.e., 257 qubits. The cost and computing power of on-demand quantum computers are $25000$ and $127$, respectively.
In the stochastic model, we consider two scenarios, i.e., $|\Omega| = 2.$ The first scenario is $\omega_1$ in which the demand of the computational task is 10, the computing power of the quantum computers is 127 qubits, and the fidelity of the Bell pairs is 1 (i.e., the best performance). The second scenario is $\omega_2$ in which there is no demand for the computational task, all quantum computers have no available computing power, and the fidelity of the Bell pairs is 0 (i.e., the worst performance). We assume the default probability values with $\pi (\omega_1)$ = 0.8 and $\pi (\omega_2) = 0.2$.

We conduct experiments via GAMS script, which can be solved by MINLP solver~\cite{gams}. Some parameters are varied in different experiments.

\subsection{Impact of Demands, Computing Power, and Fidelity}
In these experiments, we consider three important factors: i) the demand of the computational task, ii) the computing power of quantum computers, and iii) the fidelity of the Bell pairs across quantum computers.
\begin{itemize}
\item[i)] We vary the demand of the computational task from 6 to 11 qubits. Fig. \ref{fig:demand} demonstrates the cost breakdown. The quantity of deployed quantum computers also increases as the demand rises. Moreover, when the demand for computational tasks equals 11, on-demand quantum computers must be deployed to compute the computational task, as it exceeds the computing power of the currently deployed quantum computers.
\item[ii)] We vary the computing power (i.e., the number of qubits) of quantum computers. For ease of presentation, all quantum computers are identical. The cost structure, i.e., the first-stage, second-stage, and total deployment costs, is shown in Figure \ref{fig:cap}. As the computing power increases, we see that total deployment costs are falling. To achieve the lowest total cost, there is a trade-off between the first-stage cost to use the deployed quantum computers and the second-stag cost to use on-demand quantum computers.
\item[iii)] We vary the fidelity of the entangled qubits in DQC from 0 to 1. To simplify the evaluation, all the Bell pairs are identical. The results are shown in Fig. \ref{fig:fid}. When the fidelity of the Bell pairs is less than 0.5, on-demand quantum computer deployments are necessary as the entanglement of the qubits cannot be achieved. The computational tasks can be computed completely when the fidelity of the entangled qubits is equal to or greater than 0.5.
\end{itemize}

\subsection{Impact of Cost of On-Demand Quantum Computers}
We observe the total deployment cost of the proposed scheme by varying the cost of on-demand quantum computers. The cost breakdown is shown in Fig. \ref{fig:newqc}.
When the cost of an on-demand quantum computer equals to, or less than 2x, both the deployed quantum computers and on-demand quantum computers are used. 
The cost of using the deployed quantum computer is lower than those of using on-demand quantum computers when the on-demand quantum computer is more expensive, i.e., higher than 2x. Moreover, the proposed scheme can adapt the utilization of quantum computers and on-demand quantum computers to minimize the total deployment cost despite the variations in on-demand quantum computer costs.

\subsection{Impact of Probability of Scenarios}
We vary the probability of the scenario $\omega_1$, which corresponds to having the demand of the computational task, the computing power, and the positive performance of the entangled qubits. The cost breakdown is shown in Fig. \ref{fig:prob}. We observe that when the probability of the scenario is equal to or less than 0.2, the on-demand quantum computer is required due to a small amount of demand. The proposed scheme suggests using the deployed quantum computers instead of on-demand quantum computers when the probability of the demand, the computing power, and the fidelity increase.
\subsection{Cost Comparison}
We compare the total deployment cost of the proposed stochastic model with the random model and the Expected Value Formulation (EVF) model. The EVF model solves the deterministic model with using the average values of the uncertain parameters. The cost of deploying the on-demand quantum computer varies. The deployed quantum computer in the first stage is chosen at random in the random model. Figure \ref{fig:compare} compares the total costs of the three models. The proposed model, as we can see, delivers the lowest total cost. Additionally, the EVF and random models cannot adapt to the variations in costs of the on-demand quantum computers, while the proposed scheme can always achieve the most cost-effective solution.

\section{Conclusion}
In this paper, we have proposed an optimal resource allocation scheme to efficiently utilize quantum resources in DQC to compute complex computational tasks at scale. We have formulated a two-stage stochastic programming to minimize the total deployment cost under uncertainties of the demands of computational tasks, the computing power of quantum computers, and the quantum characteristics over quantum networks. The experimental findings demonstrate that the proposed scheme can minimize the total deployment cost under uncertainty.
Other methods, such as dynamic resource allocation, and other factors, such as the deployment of quantum computers across a long distance and the uncertainty of the time required to accomplish computing tasks, can be considered for future research.

\balance
\bibliographystyle{IEEEtran}
\bibliography{osdqcref}

\end{document}